\documentclass{aa}
\usepackage{times,amssymb,amsmath}
\usepackage{epsfig}

\usepackage{color}

\newcommand{\ly}{$\alpha$}

\begin{document}

\title{Narrow-band surveys for very high redshift Lyman-$\alpha$ emitters}

\author{K.K. Nilsson\inst{1,2} 
        \and A. Orsi\inst{3}
        \and C.G. Lacey\inst{4}
        \and C.M. Baugh\inst{4}
        \and E. Thommes\inst{5}
}

\institute{
   Dark Cosmology Centre, Niels Bohr Institute, Copenhagen University, Juliane 
   Maries Vej 30, 2100 Copenhagen $\O$, Denmark
\and
   European Southern Observatory, Karl-Schwarzschild-Stra{\ss}e 2, 85748 
   Garching bei M{\"u}nchen, Germany
\and
   Department of Astronomy, Pontificia Universidad Cat{\'o}lica, Casilla 306, 
   Santiago 22, Chile
\and
   Institute for Computational Cosmology, University of Durham, South Road, 
   Durham DH1 3LE, United Kingdom
\and
   Institut f{\"u}r Theoretische Physik, Universit{\"a}t Heidelberg,
   Philosophenweg 16, 69120 Heidelberg, Germany
}
\offprints{kim@dark-cosmology.dk}
\mail{kim@dark-cosmology.dk}
\date{Received date / Accepted date}

\titlerunning{Predictions for narrow-band surveys}

\abstract{
Many current and future surveys aim to detect the highest redshift 
($z \gtrsim 7$) sources through their Lyman-{\ly} (Ly$\alpha$) emission, 
using the narrow-band imaging method. However, to date the surveys have only 
yielded non-detections and upper limits as no survey has reached the necessary
combination of depth and area to detect these very young star forming
galaxies.
}
{
We aim to calculate model luminosity functions and mock surveys of
Ly$\alpha$ emitters at $z \gtrsim 7$ based on a variety of approaches
calibrated and tested on observational data at lower redshifts.
}
{
We calculate model luminosity functions at different redshifts based
on three different approaches: a semi-analytical model based on CDM, a
simple phenomenological model, and an extrapolation of observed Schechter 
functions at lower redshifts.
The results of the first two models are compared with observations made at 
redshifts $z \sim 5.7$ and $z \sim 6.5$, and they are then extrapolated to
higher redshift.
}
{
We present model luminosity functions for redshifts between
$z = 7 - 12.5$ and give specific number predictions for future planned
or possible narrow-band surveys for Ly$\alpha$ emitters. We also investigate
what constraints future observations will be able to place on the
Ly$\alpha$ luminosity function at very high redshift.
} 
{
It should be possible to observe $z = 7 - 10$ Ly$\alpha$ emitters
with present or near-future instruments if enough observing time is
allocated. In particular, large area surveys such as ELVIS (Emission
Line galaxies with VISTA Survey) will be useful in collecting a large
sample. However, to get a large enough sample to constrain well the $z
\geq 10$ Ly$\alpha$ luminosity function, instruments further in the
future, such as an ELT, will be necessary.
}


\maketitle

\section{Introduction}
One of the most promising ways of detecting very high redshift ($z
\gtrsim 5$), star-forming galaxies is via narrow-band imaging surveys
targeting Lyman-$\alpha$ (Ly$\alpha$).  In particular,
redshifts $z \sim 5.7$~and~$6.5$ have been extensively surveyed by
several groups (e.g. Ajiki et al. 2003; Hu et al. 2004; Shimasaku et
al. 2005; Ouchi et al. 2005, 2007; Malhotra et al. 2005; Taniguchi et
al. 2005; Tapken et al. 2006; Kashikawa et al. 2006).  The current
redshift record for a spectroscopically confirmed Ly$\alpha$ emitter
(LEGO -- Ly$\alpha$ Emitting Galaxy-building Object; see
M{\o}ller \& Fynbo 2001) is $z = 6.96$ (Iye et al. 2006) although
Stark et al.  (2007) have suggested the discovery of two LEGOs at $z =
8.99$~and~$9.32$.  The reason why narrow-band surveys are restricted
to a discrete number of narrow redshift windows is the night sky
OH emission lines. According to the OH line atlas of Rousselot et
al. (2000), at Ly$\alpha$ redshifts $z_{Ly\alpha} \gtrsim 7$ ($\lambda
\gtrsim 9800$~{\AA}) there are only a few possible wavelengths where a
narrow-band filter can fit in between the OH sky lines. These
correspond to $z_{Ly\alpha} \approx 7.7$,~$8.2$,~$8.8$,~$9.4$ and
$10.1 - 10.5$. Several future surveys will target these windows in the
sky aiming to detect very high redshift galaxies.  Three narrow-band
surveys for Ly$\alpha$ at redshift $z \sim 8.8$ have already been
completed (Parkes, Collins \& Joseph 1994; Willis \& Courbin 2005;
Cuby et al. 2007) but have only yielded upper limits. Future surveys
planned for these redshifts include DaZle (Dark Ages Z
Lyman-$\alpha$ Explorer, Horton et al. 2004) and ELVIS (Emission-Line
galaxies with VISTA Survey, Nilsson et al. 2006b).
Observations of very high redshift LEGOs have been proposed as
an excellent probe of reionisation, through its effects on the
Ly$\alpha$ emission line profile (e.g. Miralda-Escud{\'e} 1998;
Miralda-Escud{\'e} \& Rees 1998; Haiman 2002; Gnedin \& Prada 2004),
the luminosity function (e.g. Haiman \& Cen 2005; Dijkstra, Wyithe \&
Haiman 2007) and the clustering of sources (McQuinn et
al. 2007).

We here focus on Ly$\alpha$ emission from star-forming
galaxies, where the Ly$\alpha$ photons are emitted from gas which is
photo-ionised by massive young stars.  During recent years,
theoretical work on Ly$\alpha$ emitting galaxies has made significant
progress. There are three main aspects to these studies: \emph{i)}
predicting the numbers of star-forming galaxies as a function of star
formation rate and redshift, \emph{ii)} calculating the fraction of the
Ly$\alpha$ photons which escape from galaxies into the IGM and \emph{iii)}
calculating the factor by which the Ly$\alpha$ flux is attenuated by
scattering in the IGM on its way to the observer. Accurate treatments
of \emph{ii)} and \emph{iii)} are complicated because Ly$\alpha$ photons are
resonantly scattered by hydrogen atoms, with the consequences that
absorption of Ly$\alpha$ by dust in galaxies is hugely amplified,
thereby reducing the escape fraction, and that even a small neutral fraction
in the IGM can be effective at scattering Ly$\alpha$ photons out of
the line-of-sight, thus attenuating the flux. Because of these
complications, most theoretical papers have chosen to concentrate on
only one aspect, adopting simplified treatments of the other two
aspects.  Haiman \& Spaans (1999) made predictions of the number
counts of Ly$\alpha$ emitting galaxies by combining the
Press-Schechter formalism with a treatment of the inhomogeneous dust
distribution inside galaxies. Barton et al. (2004) and Furlanetto
et al. (2005) calculated the numbers of Ly$\alpha$ emitters in
cosmological hydrodynamical simulations of galaxy formation, but did
not directly calculate the radiative transfer of Ly$\alpha$
photons. Radiative transfer calculations of the escape of Ly$\alpha$
photons from galaxies include those of Zheng \& Miralda-Escud{\'e} (2002),
Ahn (2004) and Verhamme, Schaerer \& Maselli (2006) for idealised
geometries, and Tasitsiomi (2006) and Laursen \& Sommer-Larsen (2007)
for galaxies in cosmological hydrodynamical simulations. The
transmission of Ly$\alpha$ through the IGM has been investigated by
Miralda-Escud{\'e} (1998), Haiman (2002), Santos (2004) and Dijkstra, Lidz
\& Wyithe (2007), among others. Several authors (e.g. Haiman, Spaans \& 
Quataert 2000; Fardal et al. 2001; Furlanetto et al. 2005) have studied the 
effect of cold accretion to describe the nature of so-called Ly$\alpha$ blobs
(Steidel et al. 2000; Matsuda et al. 2004; Nilsson et al. 2006a), see also
sec.~\ref{sec:conclusion}.

Two models in particular, dissimilar in their physical assumptions,
have been shown to be successful in reproducing the observed number
counts and luminosity functions of Ly$\alpha$ emitting galaxies at
high redshifts: firstly, the phenomenological model of Thommes \&
Meisenheimer (2005)
which assumes that Ly$\alpha$ emitters are associated with the
formation phase of galaxy spheroids, and secondly the semi-analytical
model GALFORM (Cole et al. 2000, Baugh et al. 2005), which
follows the growth of structures in a hierarchical, $\Lambda$CDM
scenario. The GALFORM predictions for Ly$\alpha$ emitters are
described in detail Le Delliou et al.  (2005, 2006) and Orsi et
al. (in prep.), who show that the model is successful in reproducing
both the luminosity functions of Ly$\alpha$ emitting galaxies in the
range $3<z<6$ and also their clustering properties.

In this paper we aim to provide model predictions to help guide
the design of
future planned or possible narrow-band surveys for very high
redshift Ly$\alpha$ emitters.  We make predictions based on three
approaches: the semi-analytical and phenomenological models
already mentioned, and an extrapolation from observations at lower
redshift.  In section~\ref{sec:models} we describe the different
models used to make the predictions, and in section~\ref{sec:lf} we
present the predicted number counts and comparisons with observed
luminosity functions at lower redshifts.  In section~\ref{future} we
make number predictions for some specific future surveys. A
brief discussion regarding what can be learned from these future
surveys is found in section~\ref{constraints}. We give our
conclusions in section~\ref{sec:conclusion}.

\vskip 5mm 
Throughout this paper, we assume a cosmology with $H_0=70$
km s$^{-1}$ Mpc$^{-1}$, $\Omega _{\rm m}=0.3$ and $\Omega
_\Lambda=0.7$, apart from the mock surveys discussed in
section~\ref{constraints}, which use GALFORM models matched to the
cosmology of the Millenium Run (Springel et al. 2005), (which has 
$H_0=73$
km s$^{-1}$ Mpc$^{-1}$, $\Omega _{\rm m}=0.25$ and $\Omega
_\Lambda=0.75$).

\section{Models}\label{sec:models}
We use three different approaches to predict the numbers of high
redshift ($z>7$) Ly$\alpha$ emitters. The models
are based on very disparate assumptions.  The first model is the
semi-analytical model GALFORM (Le Delliou et al. 2005, 2006), the
second is the phenomenological model of Thommes \& Meisenheimer
(2005), and the third model is based on directly extrapolating from
observational data at lower redshifts.

Both the semi-analytical and phenomenological models assume that the
fraction of Ly$\alpha$ photons escaping from galaxies is constant, and
that the IGM is transparent to Ly$\alpha$. The simple expectation
is that before reionisation, the IGM will be highly opaque to
Ly$\alpha$, and after reionisation it will be mostly transparent. However,
various effects can modify this simple behaviour; e.g. Santos (2004)
finds that the transmitted fraction could be significant even before
reionisation, while Dijkstra, Lidz \& Wyithe (2007) argue that
attenuation could be important even after most of the IGM has been
reionised. The WMAP 3-year data on the polarisation of the microwave
background imply that reionisation occurred in the range $z \sim 8-15$
(Spergel et al. 2007), i.e. the IGM may be mostly transparent to
Ly$\alpha$ at the redshifts of most interest in this paper. In any
case, what is important for predicting fluxes of Ly$\alpha$ emitters
is the product of the escape fraction from galaxies with the
attenuation by the IGM. The two effects are in this respect
degenerate.

\subsection{Semi-analytical model}
The semi-analytical model GALFORM (Cole et al. 2000; Baugh et
al. 2005), which is based on $\Lambda$CDM, has been shown to be
successful in reproducing a range of galaxy properties at both high
and low redshift, including Ly$\alpha$ emitters in the range $z = 3 -
6$ (Le Delliou et al. 2005; 2006). A full description of GALFORM
is given in these earlier papers, so we only give a brief summary
here. GALFORM calculates the build-up of dark halos by merging, and
the assembly of the baryonic mass of galaxies through both gas cooling
in halos and galaxy mergers. It includes prescriptions for two modes
of star formation -- quiescent star formation in disks, and starbursts
triggered by galaxy mergers -- and also for feedback from supernovae
and photo-ionisation. Finally, GALFORM includes chemical evolution of
the gas and stars, and detailed stellar population synthesis to
compute the stellar continuum luminosity from each galaxy consistent
with its star formation history, IMF and metallicity (see Cole et al. 2000 
for more details). The unextincted
Ly$\alpha$ luminosity of each model galaxy is then computed from the
ionising luminosity of its stellar continuum, assuming that all
ionising photons are absorbed by neutral gas in the galaxy, with case
B recombination.

The semi-analytical approach then allows us to obtain the properties
of the Ly$\alpha$ emission of galaxies and their abundances as a
function of redshift, calculating the star formation histories for the
entire galaxy population, following a hierarchical evolution of the
galaxy host haloes. In addition, when incorporated into an N-body
simulation, we also obtain spatial clustering information. This model
has been incorporated into the largest N-body simulation to date, the
Millennium Simulation (Springel et al. 2005), to predict clustering
properties of Ly$\alpha$ galaxies. These results will be presented in
a forthcoming paper (Orsi et al., in prep.).

The version of GALFORM which we use here is the one described in
Baugh et al. (2005) and Le Delliou et al. (2006), with the same values
for parameters.  The parameters in the model were chosen in
order to match a range of properties of present-day galaxies, as well
as the numbers of Lyman Break and sub-mm galaxies at $z \sim
2-3$. We assume a Kennicutt IMF for quiescent star
formation, but a top-heavy IMF for starbursts, in order to reproduce
the numbers of sub-mm galaxies. The only parameter which has been
adjusted to match observations of Ly$\alpha$ emitters is the
Ly$\alpha$ escape fraction, which is taken to have a constant value
$f_{esc}=0.02$, regardless of galaxy dust properties. Le Delliou
et al. (2006) show that the simple choice of a constant escape
fraction $f_{esc}=0.02$ predicts luminosity functions of Ly$\alpha$
emitters in remarkably good agreement with observational data at
$3<z<6$. Le Delliou et al. (2006) also compared the predicted Ly$\alpha$
equivalent widths with observational data at $3<z<5$, including some model 
galaxies with rest-frame equivalent widths of several 100\AA, and found broad
consistency.  For this reason, we use the same value $f_{esc}=0.02$
for making most of our predictions at $z>7$. However, since the value
of the escape fraction at $z>7$ is {\em a priori} uncertain in the
models (e.g. it might increase with redshift if high redshift galaxies are
less dusty) we also present some predictions for other values of
$f_{esc}$.

Reionisation of the IGM affects predictions for the numbers of
Ly$\alpha$ emitters in deep surveys in two ways: \emph{i)} feedback from
photo-ionisation inhibits galaxy formation in low-mass halos and \emph{ii)}
reionisation changes the opacity of the IGM to Ly$\alpha$ photons
travelling to us from a distant galaxy, as discussed above. GALFORM
models the first effect in a simple way, approximating reionisation as
being instantaneous at redshift $z_{reion}$ (see Le Delliou et
al. 2006 for more details). We assume $z_{reion}=10$, in line with the
WMAP 3-year results (Spergel et al. 2007). As was shown in Le Delliou
et al. (2006; see their Fig.~8), as far as the feedback effect is concerned,
varying $z_{reion}$ over the  range $7 \lesssim z_{reion}
\lesssim 10$ does not have much effect on the bright end of the
Ly$\alpha$ luminosity function most relevant to current and planned
surveys. For example, varying $z_{reion}$ between 7 and 10 changes the
predicted luminosity function at $L_{Ly\alpha} >
10^{41.5}$~erg~s$^{-1}$ by less than 10\% for $z \sim 7-10$.

\subsection{Phenomenological model}
The phenomenological model of Thommes \& Meisenheimer (2005; TM05
hereafter) assumes that the Ly$\alpha$ emitters seen at high redshift
are galaxy spheroids seen during their formation phase. We summarise
the main features here, and refer the reader to TM05 for more
details. The model is normalised to give the observed mass function of
spheroids at $z=0$, which is combined with a phenomenological function
that gives the
distribution of spheroid formation events in mass and redshift. Each
galaxy is assumed to be visible as a Ly$\alpha$ emitter during an
initial starburst phase of fixed duration (and Gaussian in time),
during which the peak SFR is proportional to the baryonic mass and
inversely proportional to the halo collapse time. The effects of the
IMF and the escape fraction on the Ly$\alpha$ luminosity of a galaxy
are combined into a single constant factor (i.e. the escape fraction
is effectively assumed to be constant). With these assumptions, the
luminosity function of Ly$\alpha$ emitters can be computed as a
function of redshift. The free parameters in the model were chosen by
TM05 to match the observed number counts of Ly$\alpha$ emitters at
$3.5<z<5.7$ (analogously to the choice of $f_{esc}$ in the GALFORM
model). This model does not include any effects from
reionisation.

\subsection{Observational extrapolation}
Our third approach is to assume that the Ly$\alpha$ luminosity
function is a Schechter function at all redshifts, following
\begin{equation}
\phi(L)dL = \phi^{\star}(L/L^{\star})^{\alpha}\mathrm{exp}(-L/L^{\star})dL/L^{\star}
\end{equation}
and to derive the Schechter parameters $\alpha$, $\phi^{\star}$ and
$L^{\star}$ at high redshifts by extrapolating from the observed
values at lower redshifts.  For our extrapolation, we use fits to
observations at redshift $z \approx 3$ (van Breukelen et al. 2005;
Gronwall et al. 2007; Ouchi et al. 2007), $z = 3.7$ (Ouchi et
al. 2007), $z = 4.5$ (Dawson et al. 2007), $z \approx 5.7$ (Malhotra
\& Rhoads 2004; Shimasaku et al. 2006; Ouchi et al. 2007) and $z
\approx 6.5$ (Malhotra \& Rhoads 2004; Kashikawa et al. 2006), as found 
in Table~\ref{lfdata}.
\begin{table}[!t]
\begin{center}
\caption{Parameters of the fitted Schechter function in previously published
papers. References are 1) van Breukelen et al. (2005), 2) Gronwall et al. 
(2007), 3) Ouchi et al. (2007), 4) Dawson et al. (2007),
5) Malhotra \& Rhoads (2004), 6) Shimasaku et al. (2006), and 
7) Kashikawa et al. (2006). References $3-6$ 
fit for three faint end slopes 
($\alpha = -1.0, -1.5$~and~$-2.0$), but here we only reproduce the results for 
fits with $\alpha = -1.5$ as we fix the slope in our calculations. Malhotra
\& Rhoads (2004) do not give error bars on the fits. Dawson et al. (2007) 
fix the slope to $\alpha = -1.6$.}
\begin{tabular}{lccccccccc}
\hline
\hline
Ref & Redshift & $\alpha$ & $\log{\phi^{\star} \mathrm{Mpc}^{-3}}$ & $\log{L^{\star} \mathrm{ergs/s}}$ & \\
\hline
1 & $\sim 3.2$ & $-1.6$  & $-2.92^{+0.15}_{-0.23}$ & $42.70^{+0.13}_{-0.19}$ & \\
2 &       3.1  & $-1.49^{+0.45}_{-0.54}$ & $-2.84$ & $42.46^{+0.26}_{-0.15}$ & \\
3 &       3.1  & $-1.5$  & $-3.04^{+0.10}_{-0.11}$ & $42.76^{+0.06}_{-0.06}$ & \\
3 &       3.7  & $-1.5$  & $-3.47^{+0.11}_{-0.13}$ & $43.01^{+0.07}_{-0.07}$ & \\
4 &       4.5  & $-1.6$  & $-3.77^{+0.05}_{-0.05}$ & $43.04^{+0.14}_{-0.14}$ & \\
5 &       5.7  & $-1.5$  & $-4.0$ & $43.0$ & \\
6 &       5.7  & $-1.5$  & $-3.44^{+0.20}_{-0.16}$ & $43.04^{+0.12}_{-0.14}$ & \\
3 &       5.7  & $-1.5$  & $-3.11^{+0.29}_{-0.31}$ & $42.83^{+0.16}_{-0.16}$ & \\
5 &       6.5  & $-1.5$  & $-3.3$ & $42.6$ & \\
7 &       6.5  & $-1.5$  & $-2.88^{+0.24}_{-0.26}$ & $42.60^{+0.12}_{-0.10}$ & \\
\hline
\label{lfdata}
\end{tabular}
\end{center}
\end{table}
We make linear fits to $\log\phi^{\star}$ and $\log L^{\star}$ {\em
vs} $z$, and extrapolate to higher redshift.  For simplicity,
we assume a fixed faint end slope of $\alpha = -1.5$. We do not make
any corrections for any possible effects of reionisation or IGM
opacity. The extrapolated values are given in Table~\ref{lfnew}.
\begin{table}[!t]
\begin{center}
\caption{Extrapolated parameters of the observed Schechter function at higher
redshifts. The faint end slope is fixed to $\alpha = -1.5$.}
\begin{tabular}{lccccccccc}
\hline
\hline
Redshift & $\log{\phi^{\star} \mathrm{Mpc}^{-3}}$ & $\log{L^{\star} \mathrm{ergs/s}}$ & \\
\hline
7.7  & $-3.73 \pm 0.50$ & $42.88 \pm 0.24$ & \\
8.2  & $-3.80 \pm 0.50$ & $42.89 \pm 0.24$ & \\
8.8  & $-3.88 \pm 0.50$ & $42.91 \pm 0.24$ & \\
9.4  & $-3.96 \pm 0.50$ & $42.92 \pm 0.24$ & \\
12.5 & $-4.38 \pm 0.50$ & $42.99 \pm 0.24$ & \\
\hline
\label{lfnew}
\end{tabular}
\end{center}
\end{table}

\section{Luminosity functions}\label{sec:lf}
The possible Ly$\alpha$ redshifts between $z = 7$ and $z=10$ where a
narrow-band filter can be placed are $z_{Ly\alpha} = 7.7, 8.2,
8.8,$~and~$9.4$. Redshifts beyond 10 are unreachable with ground-based
instruments of the near-future. However, one possibility for $z>10$
surveys may be the James Webb Space Telescope (JWST, see
section~\ref{jwst}) and so we also make predictions for redshift
$z_{Ly\alpha} = 12.5$.  

\begin{figure}[!t]
\begin{center}
\epsfig{file=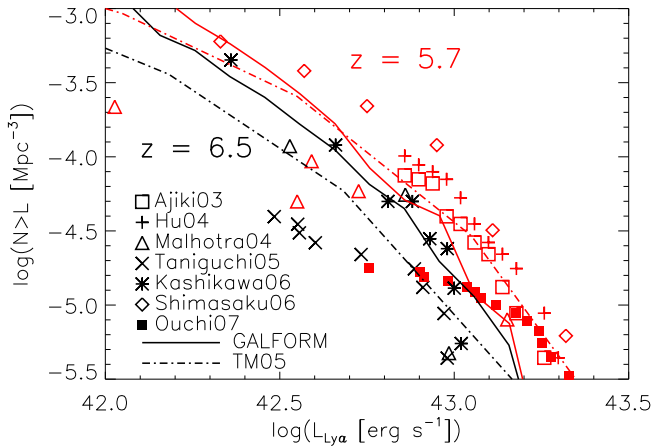,width=9.2cm}
\caption{Plot of luminosity functions at redshifts $z = 5.7$ and
$6.5$. Red points and lines are at redshift $z = 5.7$, black
points/lines at redshift $z = 6.5$. Points are observations by Ajiki
et al. (2003; redshift 5.7, squares), Hu et al. (2004; redshift 5.7,
pluses), Taniguchi et al. (2005; redshift 6.5, crosses), Shimasaku et
al. (2006; redshift 5.7, diamonds), Kashikawa et al. (2006; redshift
6.5, stars), Malhotra \& Rhoads (2004; redshift 5.7 and 6.5,
triangles) and Ouchi et al. (2007; redshift 5.7, filled
squares). Solid lines show the GALFORM model (with escape fraction
$f_{esc}=0.02$), dot-dashed lines the TM05 model. Note that the Taniguchi et 
al. (2005) and the Ouchi
et al. (2007) samples are the spectroscopic samples only.}
\label{oldplot} 
\end{center}
\end{figure}

First, we compare the Ly$\alpha$ luminosity functions predicted by the
semi-analytical (GALFORM) and phenomenological (TM05) models with
current observational data at $z \sim 6$. This comparison is shown in
Fig.~\ref{oldplot}, where we compare the models with the cumulative
luminosity functions measured in several published surveys at $z=5.7$
and $z=6.5$.
We can see that both models match the observational data reasonably
well, once one takes account of the observational uncertainties. The
error bars on the observational data points, omitted in the plot in
order to not confuse the points, are large, at the bright end of the
luminosity function due to small number statistics, and at the faint
end due to incompleteness in the samples. The shallow slopes at the
faint ends of the Taniguchi et al. (2005) and Ouchi et al. (2007)
luminosity functions may be due to spectroscopic incompleteness. Both
models fit the observations well.
Hence we conclude that both of these models can be used to extrapolate
to higher redshifts.

\begin{figure*}[!t]
\begin{center}
\epsfig{file=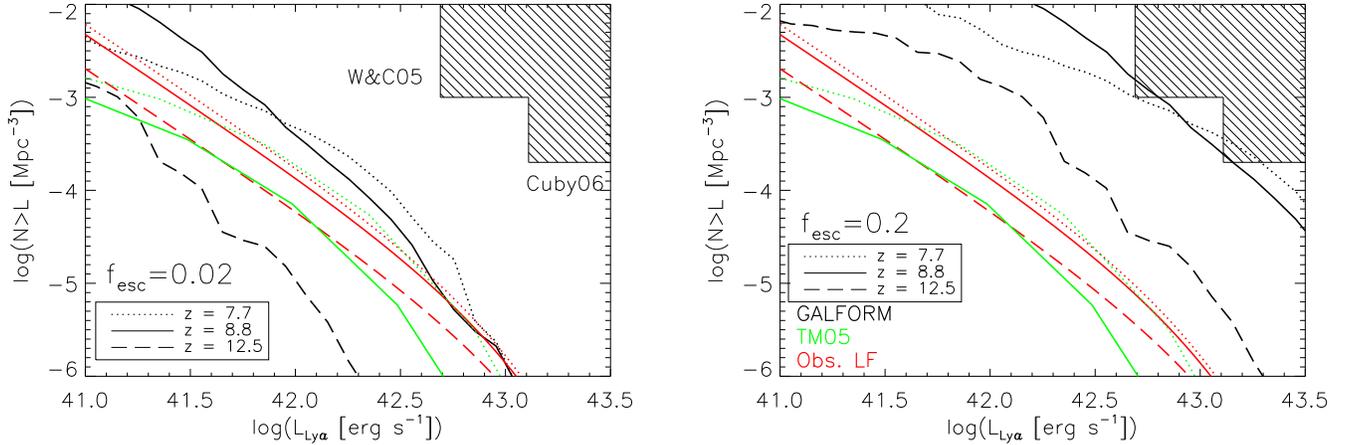,width=18cm}
\caption{Predicted Ly$\alpha$ luminosity functions at $z>7$. Red lines
are extrapolations from observed luminosity functions at lower
redshift, green lines are TM05 models and black lines are GALFORM
models. Different linestyles show different redshifts $z=7.7$, 8.8 and
12.5. No prediction is shown for the TM05 model at $z=12.5$. Hatched
area marks observational upper limits from Willis \& Courbin (2005)
and Cuby et al. (2006), both at redshift $z = 8.8$. In the left panel,
the GALFORM predictions are shown for escape fraction $f_{esc}=0.02$
(our standard value), while in the right panel, they are shown for
$f_{esc}=0.2$. The predictions from the other two methods are
identical in both panels.} 
\label{newplot} 
\end{center}
\end{figure*}

We now have three methods of extrapolating to higher redshifts,
when the direct extrapolation of the Schechter
function from lower redshifts is included.  In Fig.~\ref{newplot} we plot the
predicted luminosity functions at $z=7.7$, 8.8 and 12.5 computed by
these three methods. For other redshifts, the curves may be
interpolated. For GALFORM, we show predictions for the standard value
of the escape fraction $f_{esc}=0.02$ in the left panel, and for a
larger value $f_{esc}=0.2$ in the right panel. This illustrates the
sensitivity of the predictions to the assumed value of $f_{esc}$ at
high redshift. The predictions from the other two models are plotted
identically in both panels, since they do not explicitly include the
escape fraction as a parameter.  GALFORM predictions for the numbers
of Ly$\alpha$ emitters at $z > 7$ were also given in Le Delliou et al
(2006).
We can see that the predictions from the different methods are fairly
similar at $z = 7.7$, but gradually diverge from each other with
increasing redshift.  For the highest redshift, $z = 12.5$, the TM05
model fails in producing a prediction due to numerical problems. We
note that making predictions for $z = 12.5$ is challenging, for
several reasons. Even though only $\sim 200$~Myrs separate the ages of
the Universe between redshift 8.8~and~12.5, the Universe went through
an important transition at this time as reionisation occurred
(Spergel et al. 2007). However, we do not know exactly how and when
this happened. Also, during this epoch the structure in the dark
matter (and hence also in galaxies) was building up very rapidly. This
underlines the interest of obtaining observational constraints at
these redshifts.

The hatched regions in Fig.~\ref{newplot} show the region of
the luminosity function diagram that has been observationally excluded
at $z = 8.8$ by Willis \& Courbin (2005) and Cuby et al. (2006). The
former survey was deeper but in a smaller area, whereas the latter was
more shallow over a larger area, hence the two-step appearance of the
hatched area. From the plot, it is obvious that their non-detections
are perfectly consistent with our theoretical models, although
the GALFORM model with the non-standard escape fraction $f_{esc}=0.2$
is marginally excluded.

\section{Future surveys}\label{future}
In this  section, we discuss more specific predictions for several
planned and possible future surveys. For all calculations, we assume a
simple selection on the flux of the Ly$\alpha$ emission line, with no
additional selection on the equivalent width (i.e. we include all
galaxies with $EW_{Ly\alpha} \ge 0$). We also assume no absorption by
the neutral hydrogen in the IGM 
which would reduce the measured fluxes and for GALFORM predictions
we assume an escape fraction of $f_{esc} = 0.02$.  The predictions from the
GALFORM and TM05 models for these future surveys as well as some
published surveys are summarised in Table~\ref{tab:futnum}.

\subsection{DaZle -- Dark ages \emph{z} Lyman-{\ly} Explorer}
DaZle is a visitor mode instrument placed on the VLT UT3 (Horton et
al. 2004).  The instrument is designed to use a narrow-band
differential imaging technique, i.e. observing the same field with two
very narrow filters with slightly offset central wavelength. Objects
with Ly{\ly} in one of the filters can then be selected from the
differential image of both filters. The field-of-view of DaZle is
$6.83' \times 6.83'$ and it is expected to reach a flux level of $2
\times 10^{-18}$~erg~s$^{-1}$~cm$^{-2}$ (5$\sigma$) in 10 hours of
integration in one filter. This corresponds to a luminosity limit at
redshift $z = 7.7$ of $\log{(\mathrm{L}_{Ly\alpha})} =
42.13$~erg~s$^{-1}$.  The two initial filters are centred on
$z_{Ly\alpha} = 7.68$~and~$7.73$ (with widths $\Delta z=0.006$ and
0.025 respectively) and at this redshift, the surveyed volume becomes
$1340$~Mpc$^3$ per pointing per filter pair. Thus, from
Fig.~\ref{newplot}, we can conclude that DaZle will discover $\sim
0.16 - 0.45$ candidates at $z=7.7$ with one pointing and filter pair.

\subsection{ELVIS -- Emission Line galaxies with VISTA Survey}
ELVIS\footnote{www.astro.ku.dk/$\sim$kim/ELVIS.html} is part of
Ultra-VISTA, a future ESO Public Survey with
VISTA\footnote{www.vista.ac.uk} (Visible and Infrared Survey Telescope
for Astronomy). Ultra-VISTA is planned to do very deep near-infrared
broad- and narrow-band imaging in the COSMOS field. It will observe
four strips with a total area of 0.9~deg$^2$. The narrow-band filter
is focused on the $z_{Ly\alpha} = 8.8$ sky background window with
central wavelength $\lambda_c = 1185$~nm, and redshift width $\Delta
z=0.1$.  The flux limit of the narrow-band images is expected to reach
$3.7 \times 10^{-18}$~erg~s$^{-1}$~cm$^{-2}$ (5$\sigma$) after the
full survey has been completed. Ultra-VISTA will run from early 2008
for about 5 years and all the data will be public. ELVIS is presented
further in Nilsson et al. (2006b).  ELVIS will survey several
different emission-lines (e.g. H$\alpha$ at redshift $z = 0.8$, [OIII]
at redshift $z = 1.4$ and [OII] at redshift $z = 2.2$) as well as the
Ly{\ly} line.

When the survey is complete, the final mosaic will reach a Ly{\ly}
luminosity limit of $\log{(\mathrm{L}_{Ly\alpha})} =
42.53$~erg~s$^{-1}$. The volume surveyed will be $5.41 \times
10^5$~Mpc$^3$.  From Fig.~\ref{newplot} we see that ELVIS should be
expected to detect $3 - 20$ LEGOs at $z=8.8$.

\subsection{JWST}\label{jwst}
A possibility even further into the future is to use the James Webb
Space Telescope\footnote{www.jwst.nasa.gov} (JWST). JWST is scheduled
for launch in 2013 and will have excellent capabilities within the
near- and mid-infrared regions of the spectrum. Two of the instruments
aboard JWST could be used for narrow-band surveys; NIRCam, the
near-infrared camera, and TFI, the tunable filter imager (for a review
on JWST see Gardner et al. 2006).  NIRCam will have 31 filters, of
which nine are narrow-band filters. The filter with shortest
wavelength has central wavelength $\lambda_c = 1.644$~$\mu$m (F164N;
$z_{Ly\alpha} = 12.5$, $\Delta z = 0.135$). TFI will have tunable
filters with variable central wavelength, however it is only sensitive
at wavelengths larger than $\lambda \sim 1.6$~$\mu$m. NIRCam is
expected to reach a flux limit of $\sim 1 \times
10^{-18}$~erg~s$^{-1}$~cm$^{-2}$ (5$\sigma$) in $10000$~s of exposure
time. Hence, a flux limit of $\sim 5 \times
10^{-19}$~erg~s$^{-1}$~cm$^{-2}$ (5$\sigma$,
$\log{(\mathrm{L}_{Ly\alpha}(z = 12.5))} = 42.00$~erg~s$^{-1}$) could
be reached in 10 hours, assuming that the sensitivity is proportional
to the square root of the exposure time. TFI is expected to be able to
reach a flux limit almost a factor of two deeper in the same time,
however it has a field-of-view of only half of the NIRCam (which is $2
\times 2.16' \times 2.16'$). In one NIRCam pointing at redshift $z =
12.5$, approximately $1640$~Mpc$^3$ are surveyed. Again, from
Fig.~\ref{newplot}, we can estimate that we will detect $\lesssim 0.1$
galaxies per 10-hour pointing with NIRCam. However, the number of 
detections depends strongly on the escape fraction which is unknown at such
high redshifts, and thus the number of detected galaxies can be larger.

\section{Constraints on the early Universe}\label{constraints}
Of the surveys at these redshifts that have been presented in previous
articles (Horton et al. 2004; Willis \& Courbin 2005; Cuby et al. 2006; Nilsson
et al. 2006b), or are conceivable (JWST, see
section~\ref{jwst}) only ELVIS will detect a large enough sample to
start to measure the luminosity functions and the extent of
reionisation at these redshifts and to study the fraction of
PopIII stars in the population. We here discuss these issues with
respect to ELVIS.

\begin{figure}[!t]
\begin{center}
\epsfig{file=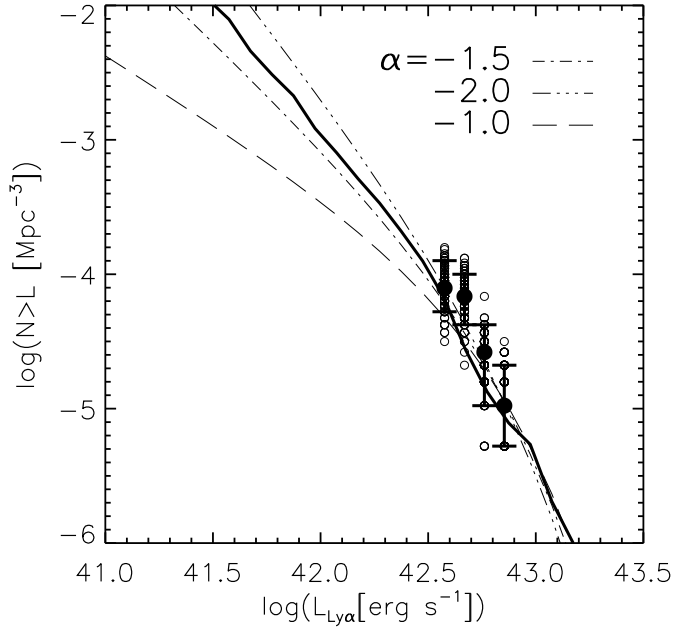,width=9.2cm}
\caption{Luminosity functions at $z=8.8$ for a set of mock ELVIS
surveys computed using GALFORM. The 112 mock surveys are identical
apart from being taken from different regions in the Millennium
simulation volume. The open circles show number counts in each mock
catalogue, in four luminosity bins. The black dot with error bars
shows the median of the mocks in each bin, with the error bars
showing the 10-90\% range. The thin lines are best fit Schechter
functions to the median points with different assumed faint end
slopes.  The thick solid line shows the ``true'' luminosity function,
as measured from model galaxies in the total Millenium simulation
volume.}
\label{cosvar} 
\end{center}
\end{figure}

From the semi-analytical modelling, we can make mock observations of
the ELVIS survey. The procedure to produce these catalogues is
explained in detail in Orsi et al. (in prep.), but the outline of the
process is that galaxies from GALFORM are placed in matching dark
matter haloes in the Millenium N-body simulation (Springel et
al. 2005), which is a cubical volume in a CDM universe of comoving
size 500~Mpc/$h$, thus creating a mock universe with simulated
galaxies which includes all the effects of clustering. We can then
make mock observations of this simulated Universe, including the
same limits on flux, redshift, sky area etc.~as for any real
survey, and from these observations produce mock galaxy
catalogues. From the mock catalogues, we can in turn make mock
luminosity functions of Ly$\alpha$ emitters at redshift $z = 8.8$. In
Fig.~\ref{cosvar} we plot the ``observed'' luminosity functions in the
112 mock catalogues taken from different regions of the Millenium
simulation volume. Note that for making these mock catalogues, GALFORM
was run with the same cosmological parameters as in the Millenium
simulation itself, which are slightly different from the
``concordance'' values assumed elsewhere in this paper, as described
in the Introduction (this is why the mean luminosity function for the
whole simulation volume which is plotted in Fig.~\ref{cosvar} is
slightly different from the GALFORM prediction for $z=8.8$ plotted in
Fig.~\ref{newplot}). We used escape fraction $f_{esc}=0.02$.
The figure shows that the spread in number density between the
different mock catalogues is large, almost a factor of ten in number
density in each luminosity bin. This is a consequence both of the
small numbers of galaxies in the mock surveys and of galaxy
clustering, which causes ``cosmic variance'' between different sample
volumes. The prediction from GALFORM is therefore that it will be
difficult to accurately measure the luminosity function of Ly$\alpha$
emitters at $z = 8.8$ even using the sample from the large area ELVIS survey. In
particular, there will be no useful constraint on the faint-end slope
$\alpha$. This is simply a consequence of the flux limit of narrow-band surveys,
i.e.~even if we use the median values of the luminosity function from
all the mocks, then Schechter functions with slopes in the range $-1$ to
$-2$ all give acceptable fits, as illustrated in
Fig.~\ref{cosvar}. However, if all the data are combined in one
luminosity bin, it should be possible to measure $\phi^{\star}$ assuming
values for $\alpha$ and $L^{\star}$. The possibility of including data points
from several surveys at different luminosities (e.g. also lensing surveys that
probe the faint end of the luminosity function) would also significantly 
improve the results.

Two suggested methods of constraining reionisation from
observations of Ly$\alpha$ emitters, without requiring spectroscopy,
are to measure the clustering of Ly$\alpha$-sources and to compare the
Ly$\alpha$ and UV continuum luminosity functions at these redshifts
(Kashikawa et al. 2006; Dijkstra, Wyithe \& Haiman 2007; McQuinn et
al. 2007). McQuinn et al. (2007) show that large HII bubbles may exist
during reionisation, and that these will enhance the observed
clustering of Ly$\alpha$ emitters in proportion to the fraction of
neutral hydrogen in the Universe. A sample of $\sim 50$ emitters will
be enough to constrain the level of reionisation using this effect
(McQuinn, priv. communication), almost within reach of the ELVIS
survey. A future, extended version of ELVIS would be able to place
very tight constraints on reionisation. In Kashikawa et
al. (2006) and Dijkstra, Wyithe \& Haiman (2007) the use of the
combination of the UV and Ly$\alpha$ LFs to constrain the IGM
transmission is explored. Ly$\alpha$ emission will be much more
susceptible to IGM absorption than the continuum emission and thus the
ratio between the two LFs will give information on the level of
IGM ionisation.  However, with increasing redshift for Ly$\alpha$, the
continuum emission will be increasingly difficult to observe, and it is
unclear if this method will be feasible for surveys such as ELVIS.

It is possible that galaxies at $z = 8.8$ still have a
significant population of primordial PopIII stars. A test for the
fraction of primordial stars is the amount of HeII~1640~{\AA} emission
(Schaerer 2003; Tumlinson, Schull \& Venkatesan 2003). Depending on
models, these authors predict that the HeII~1640~{\AA} emission line
should have a flux between $1 - 10$~\% of the flux in the Ly$\alpha$
line. For ELVIS $z = 8.8$ Ly$\alpha$ emitters, the HeII~1640~{\AA}
line is redshifted to $1.61$~$\mu$m. Due to the many OH sky emission
lines in this region of the spectrum, it would be desirable to try to
observe the HeII~1640~{\AA} line from a space-based observatory such
as JWST. According to the JWST homepage, NIRSpec will achieve a
sensitivity in the medium resolution mode on an emission line at
$1.6$~$\mu$m of $\sim 7 \times 10^{-19}$~erg~s$^{-1}$~cm$^{-2}$ ($10\sigma$) 
for an exposure time of $10^5$~s (30
hours). Thus, if the Ly$\alpha$ emission line has a flux of $\sim 5
\times 10^{-18}$~erg~s$^{-1}$~cm$^{-2}$, the HeII~1640~{\AA} will be
marginally detected with JWST in 30 hours of integration, depending on
the ratio of HeII~1640~{\AA} to Ly$\alpha$ flux. The NIRSpec
sensitivity increases at longer wavelengths, but the increasing
luminosity distance to galaxy candidates with HeII~1640~{\AA} emission
at longer wavelengths will most likely counteract this effect.

\section{Discussion}\label{sec:conclusion}
\begin{figure}[!t]
\begin{center}
\epsfig{file=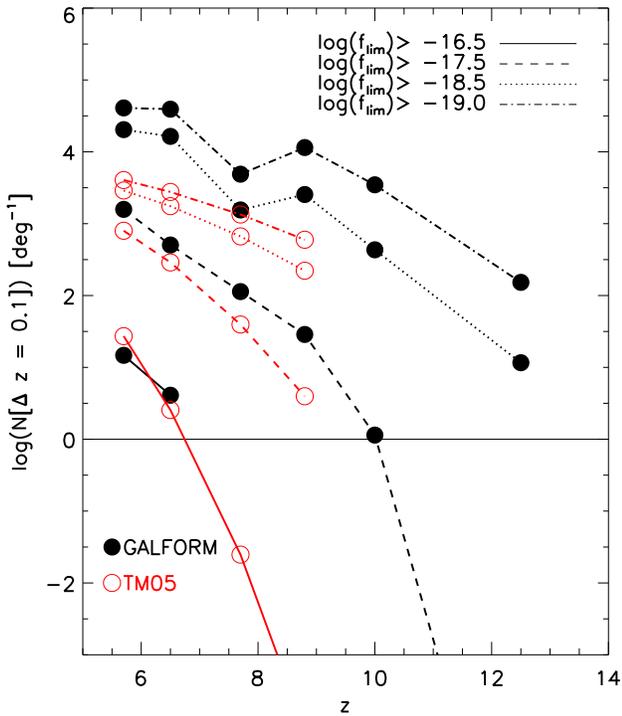,width=9.2cm}
\caption{Summary of predictions. The plot shows the number of
Ly$\alpha$ emitting galaxies expected per square degree per redshift
interval $\Delta z = 0.1$ as a function of redshift and observed flux
limit. The predictions of GALFORM are shown in black and of the TM05
model in red. The different line styles are for different flux limits.
}
\label{ngal} 
\end{center}
\end{figure}

We summarise our predictions for number counts of Ly$\alpha$ emitters
in narrow-band surveys in Fig.~\ref{ngal}.
We also summarise the numbers of detected objects for specific current
and future surveys in Table~\ref{tab:futnum}.
\begin{table*}[!t]
\begin{center}
\caption{Number of predicted/observed objects per observed field in
several present and future surveys from two theoretical models.  Data
from Subaru XMM Deep Field (SXDS) are from Ouchi et al. (2005),
Shimasaku et al. (2006) and Kashikawa et al. (2006). GALFORM 
predictions are made assuming an escape fraction of $f_{esc} = 0.02$.}
\begin{tabular}{lccccccccc}
\hline
\hline
Name & Redshift & Area (arcmin$^2$)      & Luminosity limit (5$\sigma$, erg~s$^{-1}$) & GALFORM & TM05 & Observed number &  \\
\hline
SXDS (Ouchi)       & 5.7  & 8100 & $10^{42.40}$ &  443    & 339   & 515 &  \\
SXDS (Shimasaku)   & 5.7  & 775  & $10^{42.40}$ &  112    & 86   & 83  &  \\
SXDS (Kashikawa)   & 6.5  & 918  & $10^{42.27}$ &  108    & 57   & 58   &  \\
DaZle              & 7.7  & 47   & $10^{42.13}$ &  0.45   & 0.16  & --- &  \\
ELVIS              & 8.8  & 3240 & $10^{42.50}$ &  20     & 2.8    & --- &  \\
Cuby06             & 8.8  & 31   & $10^{43.10}$ &  0.0003 & 0.0  & 0   &  \\
W\&C05             & 8.8  & 6.3  & $10^{42.25}$ &  0.015  & 0.002  & 0   &  \\
JWST               & 12.5 & 9.3  & $10^{42.00}$ &  0.018  & ---   & --- &  \\
\hline
\label{tab:futnum}
\end{tabular}
\end{center}
\end{table*}
A few comments can be made on the differences in predictions between
the two models.  Firstly, as can be seen in Fig.~\ref{ngal} and also
Fig.~\ref{newplot}, the luminosity functions have steeper faint-end
slopes in the GALFORM models than in the TM05 models. Secondly, the
GALFORM and TM05 models predict similar amounts of evolution at a
given flux over the range $z=6-9$ where they can be compared.

Several factors enter into the error bars of our predictions. One
problem is the uncertainties in, and disagreement between, the
observed lower redshift luminosity functions which are used to
calibrate the theoretical models. There are many caveats in producing
Ly$\alpha$ luminosity functions, of which the selection function is
the most difficult to correct for.  The problem arises from that the
filter transmission curve is not box-shaped, but rather
gaussian. Thus, only brighter objects will be observed at the wings of
the filter, and these will be observed to have smaller than intrinsic
luminosities. Secondly, the equivalent width (EW) limit that the
survey is complete to depends on the depth of the broad-band images
used for the selection. Thirdly, if the sample is a photometric
sample, it is possible that there are lower redshift interlopers,
where the emission line is e.g. [OII], in the sample. Finally, the
samples are still so small that we have to deal with small number
statistics.  All of these problems cause the observed luminosity
function at lower redshifts to be uncertain.

Both theoretical models (semi-analytical and phenomenological) have
uncertainties resulting from how they model the galaxy formation
process, and also from the assumption that the fraction of Ly$\alpha$
photons escaping from galaxies is constant and does not change with
redshift. In addition, neither model includes attenuation of the
Ly$\alpha$ flux due to neutral hydrogen in the IGM. This attenuation
would be expected to be strong at $z>z_{reion}$, when the IGM is
neutral, and weaker at $z<z_{reion}$, when most of the IGM is
ionised. The degree of attenuation depends on a number of different
effects, as analysed in Santos (2004), and discussed in Le Delliou et
al. (2006), and is currently very uncertain. Nonetheless, this
attenuation is expected to produce observable effects on the evolution
of the Ly$\alpha$ luminosity function, if reionisation occurs within
the redshift range covered by future observations, and so estimating
the reionisation redshift and the neutral fraction after reionisation
are included in the science goals of these surveys.

It is apparent that the key to acquiring a large sample of
Ly$\alpha$-emitting galaxies at redshifts greater than 7 is both depth
and area. In a recent paper, Stark, Loeb \& Ellis (2007) suggest that
one of the most efficient means of finding very high redshift
Ly$\alpha$ emitters is through spectroscopic surveys focused on
gravitational lensing clusters. Lensing surveys could easily reach
down to a luminosity limit of $10^{40.5}$~erg~s$^{-1}$ in a few tens
of hours. However, the surveyed volumes are very small, of the order
of a hundred Mpc$^3$. For a lensed survey, the area in the source
plane is reduced by the same factor that the flux is amplified, so in
principle one gains in the total number of objects detected relative
to an unlensed survey if the luminosity function is steeper than
$N(>L) \propto L^{-1}$. In the GALFORM and TM05 models, the asymptotic
faint-end slope is shallower than this, but at higher luminosities,
the slope can be steeper. For example, GALFORM predicts that at
$z=10$, the average slope in the luminosity range
$10^{41}$--$10^{42}$~erg~s$^{-1}$ is close to $N(>L) \propto L^{-2}$
(see Fig.~8 in Le~Delliou et al. 2006), so that a lensing amplification
of 10 results in 10 times more objects being detected, with intrinsic
luminosities 10 times lower, compared to an unlensed survey with the
same area and flux limit.
Therefore lensing and narrow-band surveys are complementary to each
other as they probe different parts of the luminosity function.  With
either type of survey, reaching a significant sample of redshift $z
\sim 7 - 8$ should be possible in the next few years with
telescopes/instruments in use or soon available.

An interesting type of object found recently in narrow-band
surveys are the Ly$\alpha$ blobs, large nebulae with diameters up to
150 kpc and Ly$\alpha$ luminosities up to $10^{44}$~erg~s$^{-1}$ with
or without counterpart galaxies (e.g. Steidel et al. 2000; Matsuda et
al. 2004; Nilsson et al. 2006a). Several mechanisms have been proposed
to explain this phenomenon, including starburst galaxies and
superwinds, AGN activity or cold accretion. It is interesting to
consider if such objects would be detected in any of these surveys,
assuming they exist at these redshifts.  A typical Ly$\alpha$ blob
will have a luminosity of $\sim 10^{43}$~erg~s$^{-1}$ and a radius of,
say, 25~kpc. This will result in a surface brightness of $\sim 5
\times 10^{39}$~erg~s$^{-1}$~kpc$^{-2}$. Thus, a narrow-band survey
will have to reach a flux limit, as measured in a $2''$~radius
aperture of $\sim 1.3 \times 10^{42}$~erg~s$^{-1}$ at redshift $z =
8.8$, corresponding to $\log{L} = 42.11$. (An aperture radius of
around $2''$ is expected to be roughly optimal for signal-to-noise.)
For lower or higher redshifts, this limit is higher or lower
respectively. Thus, ELVIS will not be able to detect Ly$\alpha$ blobs
unless they are brighter and/or more compact at higher redshift than a
typical blob at lower redshift. DaZle and JWST could in principle
detect this type of object, but only if they are very abundant in the
very high redshift Universe, due to the small survey volumes of these
instruments. It is of course highly uncertain what properties such
Ly$\alpha$ blobs would have at $z \sim 7 - 9$, or their space density,
but it appears unlikely that the future surveys presented here would
detect any such objects.

To find compact Ly$\alpha$ emitters at redshifts $z \gtrsim 10$ in
significant numbers we will probably have to await instruments even
further in the future. If a future 40-m ELT (Extremely Large
Telescope) was equipped with a wide-field NIR imager and a
narrow-band filter of similar width to ELVIS, it could reach a
luminosity limit of $L \sim 10^{41.2}$~erg~s$^{-1}$ at redshift $z =
10.1$ (where a suitably large atmospheric window exists) in
approximately 20 hours. Using the GALFORM model for $z = 10$, the
number density should be N($>$L)$\approx 4 \times 10^{-3}$~Mpc$^{-3}$
at this luminosity limit. Thus, to get a sample of ten Ly$\alpha$
emitters would require imaging an area on the sky of approximately
16~square arcminutes, assuming a narrow-band filter with
redshift range $10.05 < z_{\mathrm{Ly}\alpha} < 10.15$. This could be
achieved with one pointing if the detector has a field-of-view of
6~arcmin on a side, as suggested by the ESO ELT Working
Group\footnote{http://www.eso.org/projects/e-elt/Publications/ELT\_INSWG\_FINAL\_REPORT.pdf}. It
should of course be noted that these are very tentative numbers, but
they display the possibilities of far future instruments.

\begin{acknowledgements}
The authors wish to thank J. Fynbo, P. M{\o}ller, O. M{\"o}ller,
J. Sommer-Larsen and Jean-Paul Kneib for comments on the manuscript. KN 
gratefully
acknowledges support from IDA~--~Instrumentcenter for Danish
Astrophysics. The Dark Cosmology Centre is funded by the DNRF.  AO
acknowledges support of the European Commission's ALFA-II programme
through its funding of the Latin-american European Network for
Astrophysics and Cosmology (LENAC). CGL is supported by the PPARC
rolling grant for extragalactic astronomy and cosmology at Durham. CMB
is supported by a Royal Society University Research Fellowship.

\end{acknowledgements}


\begin{thebibliography}{99}
\bibitem{Ahn}Ahn, S.-H., 2004, ApJ, 601, L25
\bibitem{Aji03}Ajiki, M., Taniguchi, Y., Fujita, S.S., et al., 2003, AJ, 126, 2091
\bibitem{barton}Barton, E.J., Dav{\'e}, R.,; Smith, J.-D.T., et al., 2004, ApJ, 604, L1
\bibitem{baugh}Baugh, C.M., Lacey, C.G., Frenk, C.S., et al., 2005, MNRAS, 356, 1191
\bibitem{cole}Cole, S., Lacey, C.G., Baugh C.M., \& Frenk, C.S., 2000, MNRAS, 319, 168
\bibitem{Cuby}Cuby, J.-G., Hibon, P., Lidman, C., et al., 2007, A\&A, 461, 911
\bibitem{Daw}Dawson, S., Rhoads, J.E., Malhotra, S., et al., 2007, submitted to ApJ, arXiv:0707.4182
\bibitem{Del05}Le Delliou, M., Lacey, C.G., Baugh, C.M., et al., 2005, MNRAS, 357, L11
\bibitem{Del06}Le Delliou, M., Lacey, C.G., Baugh, C.M., \& Morris, S.L., 2006, MNRAS, 365, 712
\bibitem{Dijk}Dijkstra, M., Wyithe, S., \& Haiman, Z., 2007, MNRAS, 379, 253
\bibitem{Dijk2}Dijkstra, M., Lidz, A., \& Wyithe, J.S.B., 2007, MNRAS, 377, 1175
\bibitem{Fardal}Fardal, M.A., Katz, N., Gardner, J.P. et al., 2001, ApJ, 562, 605
\bibitem{furla}Furlanetto, S.R., Schaye, J., Springel, V., \& Hernquist, L., 2005, ApJ, 622, 7
\bibitem{gardner}Gardner, J.P., Mather, J.C., Clampin, M., et al., 2006, Space Science Reviews, 123, 485, astro-ph/0606175 
\bibitem{Gne}Gnedin, N.Y. \& Prada, F., 2004, ApJ, 608, L77
\bibitem{gronwall}Gronwall, C., Ciardullo, R., Hickey, T., et al., 2007, accepted for publication in ApJ, arXiv:0705.3917
\bibitem{Hai1}Haiman, Z. \& Spaans, M., 1999, ApJ, 518, 138
\bibitem{Haiman2}Haiman, Z., Spaans, M. \& Quataert, E., 2000, ApJL 537, L5
\bibitem{Hai2}Haiman, Z., 2002, ApJ, 576, L1
\bibitem{Hai3}Haiman, Z. \& Cen, R., 2005, ApJ, 623, 627
\bibitem{Hor}Horton, A., Parry, I., Bland-Hawthorne, J., et al., 2004, astro-ph/0409080
\bibitem{Hu2}Hu, E.M., Cowie, L.L., Capak, P., et al., 2004, ApJ, 127, 563
\bibitem{Iye}Iye, M., Ota, K., Kashikawa, N. et al. 2006, Nature, 443, 186
\bibitem{Kashi}Kashikawa, N., Shimasaku, K., Malkan, M.A., et al., 2006, ApJ, 648, 7
\bibitem{pela}Laursen, P., \& Sommer-Larsen, J., 2007, ApJ, 657, L69
\bibitem{Mal2}Malhotra, S. \& Rhoads, J., 2004, ApJ, 617, L5
\bibitem{Mal3}Malhotra, S., Rhoads, J.E., Pirzkal, N., et al., 2005, ApJ, 626, 666
\bibitem{Matsuda}Matsuda Y., Yamada T., Hayashino T. et al., 2004, AJ, 128, 569
\bibitem{McQ}McQuinn, M., Hernquist, L., Zaldarriaga, M., \& Dutta, S., 2007, accepted by MNRAS, astro-ph/0704.2239
\bibitem{Mir1}Miralda-Escud{\'e}, J., 1998, ApJ, 501, 15
\bibitem{Mir2}Miralda-Escud{\'e}, J. \& Rees, M.J., 1998, ApJ, 497, 21
\bibitem{Moeller}M{\o}ller, P., \& Fynbo, J.U., 2001, A\&A, 372, L57 
\bibitem{Nil}Nilsson, K.K., Fynbo, J.P.U., M{\o}ller, P., Sommer-Larsen, J., \& Ledoux, C., 2006a, A\&A, 452, L23
\bibitem{Nil2}Nilsson, K.K., Fynbo, J.P.U., M{\o}ller, P., \& Orsi, A., 2006b, to appear in the ASP Conference proceedings of 'At the Edge of the Universe', eds. J. Afonso, H. Ferguson and R. Norris, astro-ph/0611239
\bibitem{Ouc}Ouchi, M., Shimasaku, K., Akyiama, M. et al., 2005, ApJ, 620, L1
\bibitem{Ouc2}Ouchi, M., Shimasaku, K., Akiyama, M., et al., 2007, submitted to ApJ, arXiv:0707.3161
\bibitem{Park}Parkes, I.M., Collins, C.A., \& Joseph, R.D., 1994, MNRAS, 266, 983
\bibitem{rousselot}Rousselot, P., Lidman, C., Cuby, J.-G., Moreels, G., \& Monnet, G., 2000, A\&A, 354, 1134
\bibitem{Santos}Santos, M.R., 2004, MNRAS, 349, 1137
\bibitem{Schaerer}Schaerer, D., 2003, A\&A, 397, 527
\bibitem{shima}Shimasaku, K., Kashikawa, N., Doi, M., et al., 2006, PASJ, 58, 313
\bibitem{spergel}Spergel, D.N., Bean, R., Dor{\'e}, O, et al., 2007,
  ApJS, 170, 377
\bibitem{springel}Springel, V., White, S.D.M., Jenkins, A., et al., 2005, Nature, 435, 629
\bibitem{stark}Stark, D.P., Ellis, R.S., Richard, J., et al., 2007, ApJ, 663, 10
\bibitem{stark2}Stark, D.P., Loeb, A., \& Ellis, R.S., 2007, submitted to ApJ, astro-ph/0701882
\bibitem{Ste}Steidel, C.C., Adelberger, K.L., Shapley, A.E, et al., 2000, ApJ, 532, 170
\bibitem{Tan}Taniguchi, Y., Ajiki, M., Nagao, T. et al., 2005, PASJ, 57, 165
\bibitem{Tap}Tapken, C., Appenzeller, I., Gabasch, A., et al., 2006, A\&A, 455, 145
\bibitem{tasi}Tasitsiomi, A., 2006, ApJ, 645, 792
\bibitem{Tho}Thommes, E. \& Meisenheimer, K., 2005, A\&A, 430, 877
\bibitem{Tuml}Tumlinson, J., Shull, J.M., \& Venkatesan, A., 2003, ApJ, 584, 608
\bibitem{vanB}van Breukelen, C., Jarvis, M. J., \& Venemans, B. P. 2005, MNRAS, 359, 895
\bibitem{Verhamme}Verhamme, A., Schaerer, D., \& Maselli, A., 2006, A\&A, 460, 397
\bibitem{Wil}Willis, J.P. \& Courbin, F., 2005, MNRAS, 357, 1348
\bibitem{Zheng} Zheng, Z., \& Miralda-Escud{\'e}, J.  2002, ApJ 578,33
\end{thebibliography}
\end{document}